\documentclass[apjl]{emulateapj}

\def\pdeg{\ifmmode $\setbox0=\hbox{$^{\circ}$}\rlap{\hskip.11\wd0 .}$^{\circ}
          \else \setbox0=\hbox{$^{\circ}$}\rlap{\hskip.11\wd0 .}$^{\circ}$\fi}
\def\arcs{\ifmmode {^{\scriptscriptstyle\prime\prime}}
          \else $^{\scriptscriptstyle\prime\prime}$\fi}
\def\arcm{\ifmmode {^{\scriptscriptstyle\prime}}
          \else $^{\scriptscriptstyle\prime}$\fi}

\begin{document}

\title{Discovery of a 500~pc shell in the nucleus of Centaurus A}

%\affil{JPL}
\author{Alice C. Quillen\altaffilmark{1}, 
Joss Bland-Hawthorn\altaffilmark{2},
Mairi H. Brookes\altaffilmark{3},  
Michael W. Werner\altaffilmark{3},
J. D. Smith\altaffilmark{4},
Daniel Stern\altaffilmark{3},
Jocelyn Keene\altaffilmark{3},
Charles R. Lawrence\altaffilmark{3}
} 
\email{aquillen@pas.rochester.edu}
\email{jbh@aao.gov.au} 
\altaffiltext{1}{Department of Physics and Astronomy, University of Rochester, Rochester, NY 14627}
\altaffiltext{2}{Anglo-Australian Observatory, P.O. Box 296, Epping NSW, Australia} 
\altaffiltext{3}{Jet Propulsion Laboratory, 4800 Oak Grove Drive, Pasadena, CA 91109}
\altaffiltext{4}{Steward Observatory, University of Arizona, 933 North Cherry Avenue, Tucson, AZ 85721}
% \email{mbrookes@mail.jpl.nasa.gov}
% \email{mwerner@sirtfweb.jpl.nasa.gov}
%\email{jdsmith@as.arizona.edu}
%\author{Jocelyn Keene}    \email{jkeene@sirtfweb.jpl.nasa.gov} 
%\author{Charles R. Lawrence} \email{clawrence@sirtfweb.jpl.nasa.gov}
%\author{Daniel Stern}     \email{stern@zwolfkinder.jpl.nasa.gov}
%\author{Peter R. Eisenhardt} \email{peisenhardt@sirtfweb.jpl.nasa.gov}
%\author{Karl R. Stapelfeldt} \email{krs@ipac.caltech.edu}
%\author{Varoujan Gorjian} \email{vg@jpl.nasa.gov}
%\author{...  order and list to be determined}
%\author{B\"arbel Koribalski} \email{Baerbel.Koribalski@atnf.csiro.au}
%\affil{Australia Telescope National Facility, CSIRO, PO Box 76, Epping, NSW 1710, Australia}

\begin{abstract}

{\it Spitzer Space Telescope} mid-infrared images of the radio galaxy 
Centaurus A
reveal a shell-like, bipolar, structure 500~pc to the north and south
of the nucleus. This shell is
seen in 5.8, 8.0 and 24$\mu$m broad-band images. 
Such a remarkable shell has not been previously detected in 
a radio galaxy and is the first  
extragalactic nuclear shell detected at mid-infrared wavelengths.
We estimate that the shell
is a few million years old and has a 
mass of order million solar masses.  A conservative estimate
for the mechanical energy in the wind driven bubble 
is $10^{53}$ erg.  
The shell could have created by a small  few thousand solar mass 
nuclear burst of star formation. 
Alternatively, the bolometric
luminosity of the active nucleus is sufficiently large that
it could power the shell.
Constraints on the shell's velocity are lacking. 
However, if the shell is moving at 1000~km~s$^{-1}$
then the required mechanical energy would be 100 times larger. 

\end{abstract}


\section{Introduction}

The nearest of all the giant radio galaxies, Centaurus~A 
(NGC~5128) provides a unique opportunity to observe
the dynamics and morphology of an active galaxy 
in detail across the electromagnetic spectrum.
For a recent review
on this remarkable object see \citet{israel}.  
In its central regions, NGC~5128 exhibits
a well recognized, optically-dark band of absorption
across its nucleus. 
{\it Spitzer} images of the galaxy reveal a parallelogram shape 
\citep{quillen_irwarp}
that has been modeled as a series of folds
in a warped thin disk (e.g., 
\citealt{bland86,bland87,nicholson92,sparke96,quillen_irwarp}).

In this letter we report on the discovery of a 500~pc sized 
bipolar shell in mid-infrared images in the center of the galaxy.
Shells have been previously seen in active and star forming galaxies 
(for a recent review on galactic winds see \citealt{veilleux05}).
For example,
the Seyfert 2 galaxy NGC~2992 exhibits a figure-eight
shaped morphology in the radio continuum \citep{ulvestad84}.
% although the narrow-line morphology is
%dominated by an ionization bi-cone apparently unrelated to
%the radio continuum lobes (Allen et al. 1999).
The starburst/LINER galaxies NGC~2782 \citep{jogee98},
and NGC~3079 \citep{ford86,veilleux94}, 
%Cecil et al. 1999, Veilleux 2000)
exhibit partially closed shell morphologies in 
optical emission lines and radio continuum. 
%From HI observations,
%\citet{mcclure00} discovered $\sim 500$~pc sized shells 
%in the outer Milky Way. 

The above shells have been detected in radio continuum, optical
emission lines and in emission from atomic hydrogen.  
The only nuclear shell to have been previously detected at
mid-infrared wavelengths is the 170~pc large
bipolar bubble in the direction of the Galactic center 
%detected with {\it Midcourse Space Experiment} observations 
\citep{bland03}.
As pointed out by \citet{bland03}, infrared observations
allow unique constraints on the mass of swept up material.
Theoretical models predict that in the early stages of
a starburst driven outflow supernovae and stellar winds inject
energy into the ISM forming a bubble of hot gas and thermalized ejecta 
(e.g., \citealt{castor75,chevalier85,tomisaka88,heckman90}).
Active galaxies can also drive winds 
(e.g., see \citealt{veilleux05,begelman04}).

Based on the discussion by \citet{israel},
we adopt a distance to Cen A of $3.4$~Mpc.
At this distance $1'$ on the sky corresponds to $\sim 1$~kpc.

%section 2
\section{Observations}

\begin{figure*}
\epsscale{0.8}
\plotone{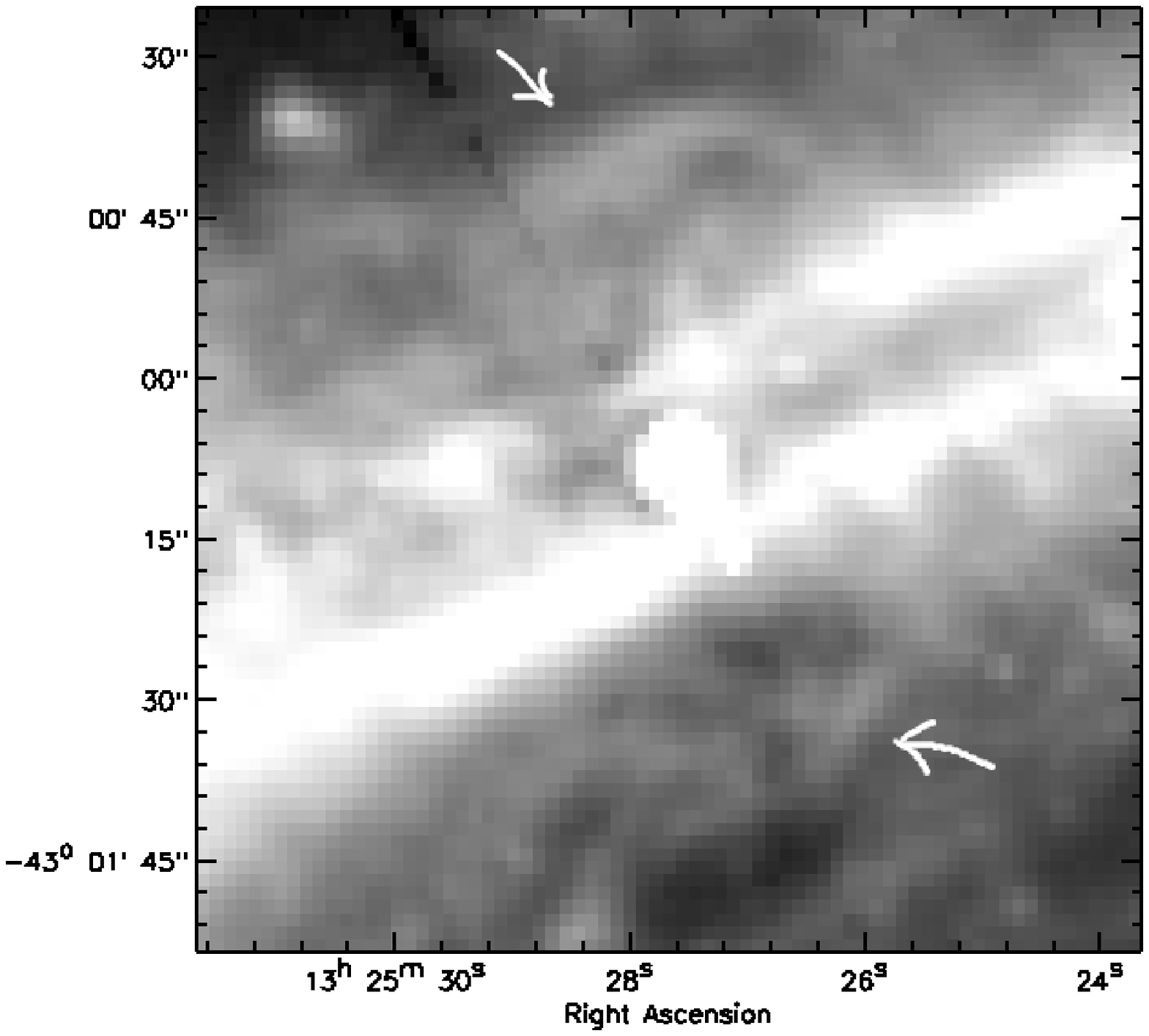}
\plotone{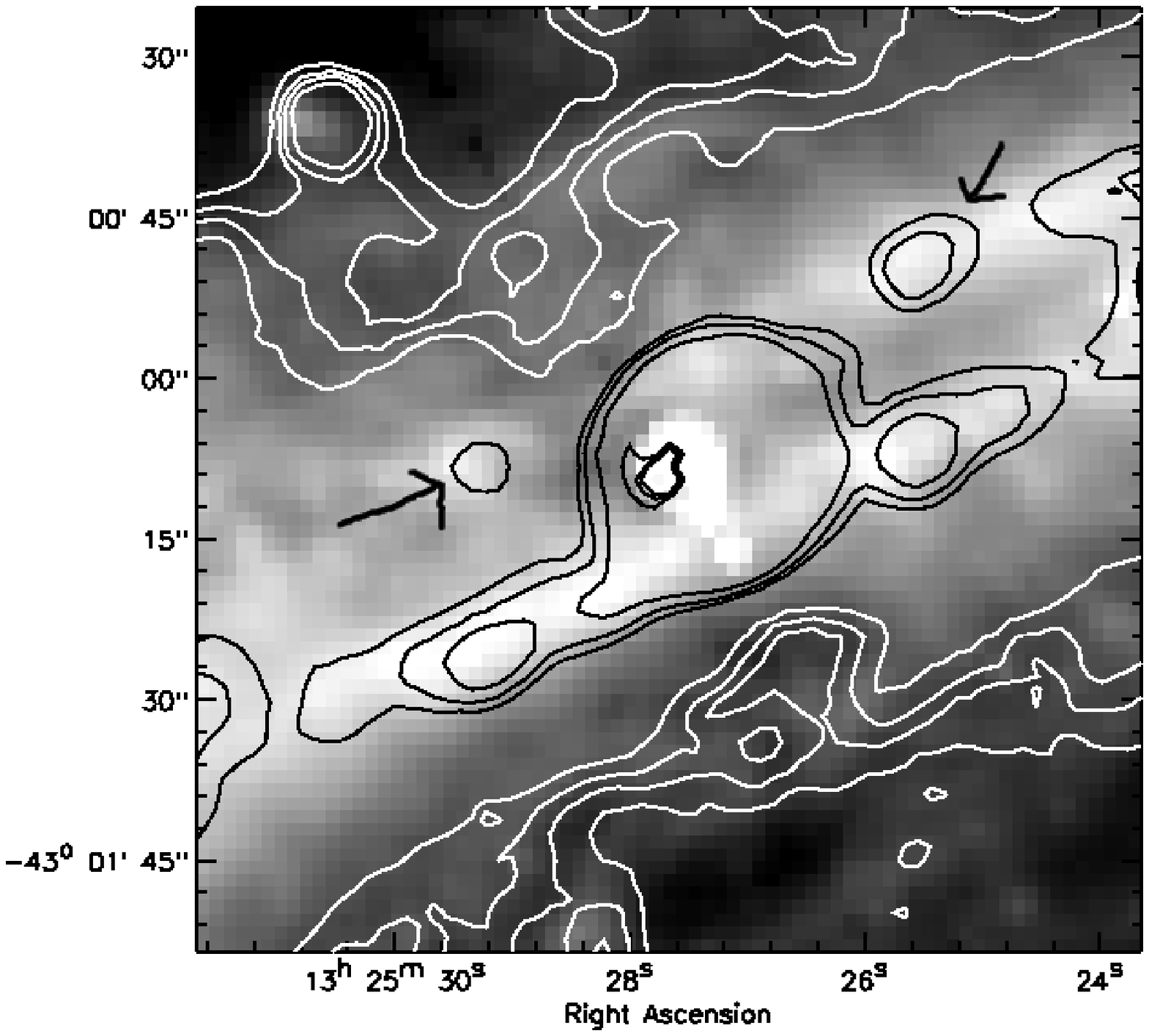}
\figcaption{
a) Nuclear region showing the bipolar shell at 
8.0$\mu$m.  The emission is shown on a log scale.
The 8.0$\mu$m emission has been subtracted by that
at $3.6\mu$m, removing some of the contribution from the background 
starlight.
The rims of the shell are emphasized with white arrows.
The shell rims are $30\arcs$ or 500~pc from the galaxy nucleus.
b) The $8.0\mu$m image (grayscale) overlayed with contours from the 
{MIPS} 24$\mu$m image (again with the $3.6\mu$m image subtracted).
Contours are shown at 
2, 4, 6, 9, 56, 70, and 94~MJy~sr$^{-1}$ above the sky background.
The shell is evident at 5.8, 8.0, and $24\mu$m.
We point out with black arrows 
the upper two of the four bright points discussed
in the text in section 2.
\label{fig:nucimages}
}
\end{figure*}

In Fig.~\ref{fig:nucimages} we show 8.0 and 24$\mu$m broad-band images
from the Infrared Array Camera ({IRAC}) 
and Multiband Imaging Photometer for {\it Spitzer}  (MIPS)
on board the {\it Spitzer Space Telescope}.  
These images have been described 
by \citet{quillen_irwarp,brookes06}.
An oval or bipolar shell is evident in these images $30\arcs$ both above
and below the nucleus.  
Along the minor axis of the shell,
four bright spots lie in the parallelogram feature in the {MIPS} 
$24\mu$m image (see the dark contours in Fig.~\ref{fig:nucimages}b);
these are also seen in the $8\mu$m image.  
The warped disk models, \citep{quillen_irwarp},
show that the parallelogram shape
is caused by folds in the disk located at a radius of $\gtrsim 60\arcs$,
and outside the expected location of the shell, at half of this radius.  
Because of the difference in estimated radius, 
the four bright points  
are unlikely to be due to an interaction between the shell 
and the warped disk.
They are probably due to a superposition of the shell (along
its minor axis) and the warped disk.

%par2
We estimate that the shell has length 1.1' and width 0.7' corresponding to
an axis ratio of $\sim 0.63$.
At the distance of the galaxy, the angular scale of the shell
places the rims at a distance of $\sim 500$~pc from the galaxy nucleus
along the shell's major axis, north and south of the nucleus.
The shell is oriented with major axis at 
a position angle $PA \sim 10^\circ$. % counter-clockwise from north.
%This position angle can be compared to the jet axis in the same region.  
The radio jet is seen to the north-east of the 
nucleus at $PA \sim 55^\circ$ \citep{burns83}; this 
differs significantly from the shell's major axis.  
The galaxy isophotes at large radii, $r > 4'$,  are elongated
approximately along the same axis as that of the shell.
However, inside a radius of $2'$ the underlying elliptical 
galaxy component is nearly spherical  (see isophotes shown by 
\citealt{quillen_irwarp}).  More relevant is that the 
gaseous and dusty disk is approximately perpendicular to
the shell.  The shell orientation would be consistent with a bipolar bubble
expanding above and below the more massive and denser
stellar and gaseous disk.

%par3
Because the shell features are detected in more than
one band ({IRAC} bands 3 and 4) and in more than one camera 
(both {IRAC}
and {MIPS}), the feature is very likely to be real, and not an
artifact of the {\it Spitzer Space Telescope} or due to scattered
light from the nucleus.  No artifact with similar structure
has been reported by the instrument teams 
from observations of bright sources
(e.g., \citealt{fazio,rieke}).

%par4
A cut along the major axis of the shell
at $PA=10^\circ$ is shown as a surface brightness
profile in Fig.~\ref{fig:slice}. 
The shell can be seen as bumps at a distance $\pm 30\arcs$ from
the nucleus.  By subtracting a linear 
fit to the background emission to either side of the shell,
we measure peak surface brightnesses in the shell of 2.5, 10, and 
10~MJy~sr$^{-1}$ at 5.8, 8.0 and 24$\mu$m, respectively.
These measurements are approximate (uncertain by a factor of 50\%) due to 
the uncertainty in the fit to the background.
The flux ratios between the bands are not untypical 
of emission from dust in nearby galaxies; as compared
to those listed by \citet{dale05}.

\begin{figure*}
\plotone{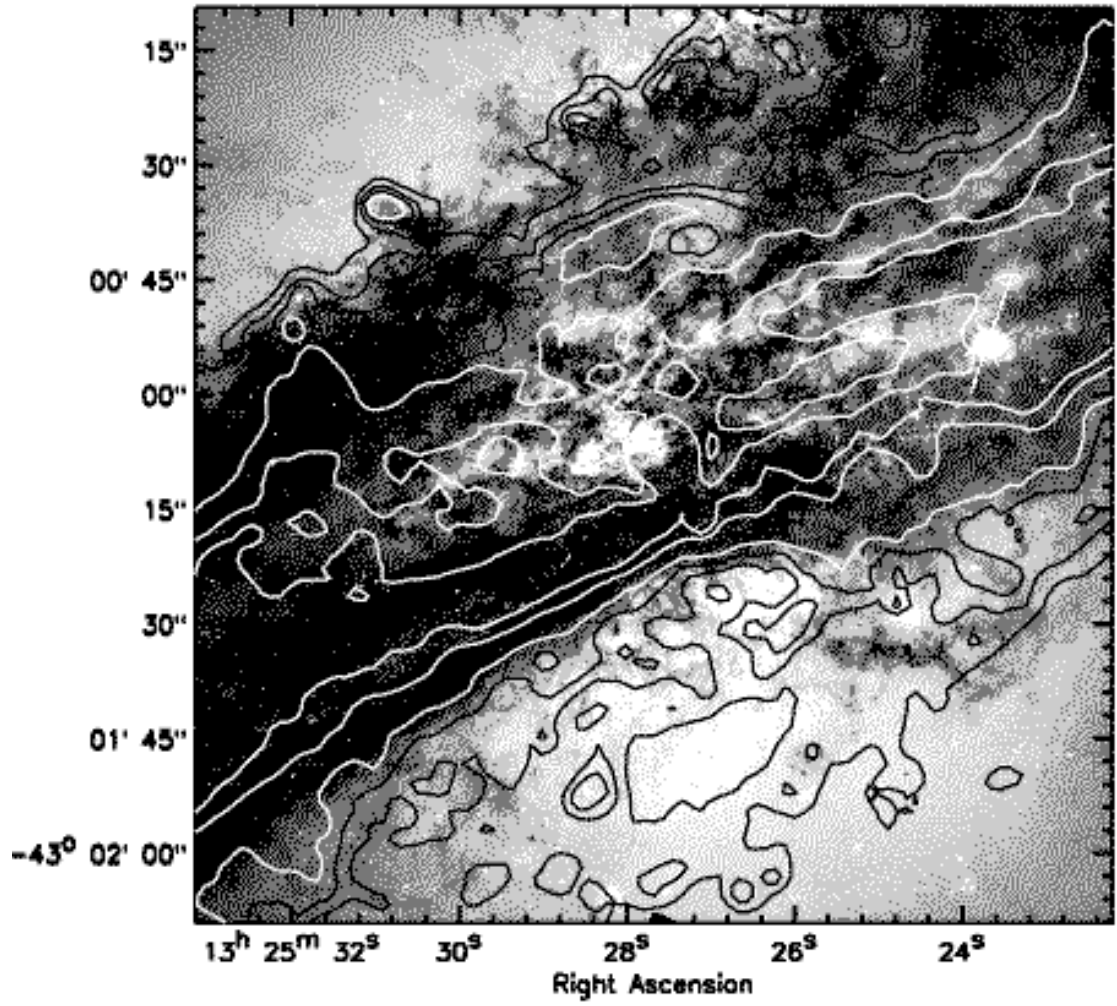}
\figcaption{
a) A somewhat larger view of the same region of the nucleus 
viewed in grayscale at $0.55\mu$m  
taken with the {WFPC2} camera on board {\it HST} (see \citealt{marconi00})
overlayed with contours from the IRAC $8.0\mu$m image.
Contours are shown at 13, 18, 23, 30, 56 and 90~MJy~sr$^{-1}$.
Extinction features at $0.55\mu$m are coincident with the infrared
shell rim $30\arcs$ south of the nucleus.  There may be a larger shell 
$45\arcs$ south of the nucleus evident in the lower contour 
at $8.0\mu$m.
%Shown is an image at $0.55\mu$m in the F555W broad-band filter.
%The {\it HST} image has been previously discussed by \citet{marconi00}.
%b) color map made from a division of the 
%band 2 ($4.5\mu$m) and band 3 ($5.8\mu$m) {IRAC} images.  
\label{fig:wfpc}
}
\end{figure*}

\begin{figure*}
\epsscale{0.6}
\plotone{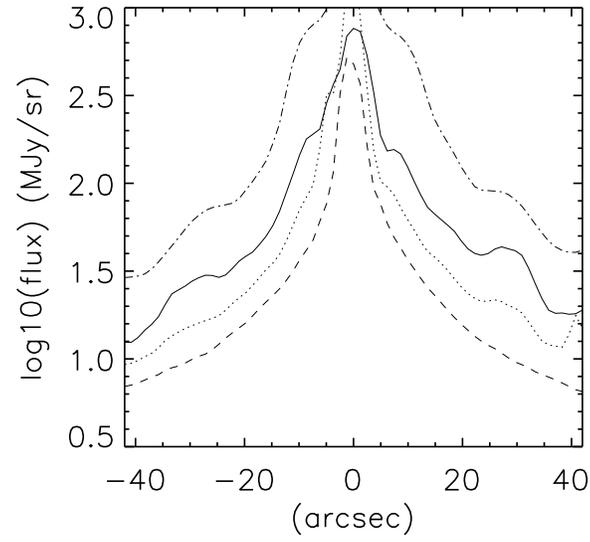}
\figcaption{
Surface brightness as a function of distance
from the nucleus in a $3\arcs$ wide slice oriented along
$PA = 10^\circ$ containing and centered at the galaxy nucleus.  
The right hand side corresponds to positions north
of the nucleus.  From top to bottom lines show the log$_{10}$ 
of the surface brightness at $24\mu$m,  offset upward by $+0.5$, that at
8.0, 5.8 and 4.5$\mu$m, respectively, and with no offsets.
The shell is visible as bumps at $\pm 30\arcs$ from the nucleus.
\label{fig:slice}
}
\end{figure*}

%{\bf Comparison to other data:}
%par5
No prominent feature coincident with the shell is clearly seen 
in the imaging  done with the {\it Hubble Space Telescope}
({\it HST}) (e.g., \citealt{schreier,marconi00}), 
though the filaments south of the nucleus are coincident with
those in the shell in the 8.0 and 5.8$\mu$m {IRAC} images 
(see Fig.~\ref{fig:wfpc}).
Further than $10\arcs$ south of the nucleus 
the galaxy center is not obscured by folds of the gaseous and dusty disk so
a dusty structure could
be better seen in absorption against the underlying galaxy 
(see \citealt{quillen_irwarp} for further discussion on the morphology
of the warped disk).  
North of the nucleus, a fold of the disk, evident as
a dark band across the {\it HST} image,  obscures the 
region where the shell is located.  
This may account for our failure
to match absorption features in the visible {\it HST} images with 
the infrared shell on the northern side, but our ability
to do so on the southern side.
The dust extinction features in the visible broad-band Wide Field
Planetary Camera-2 (WFPC2) {\it HST} images
appear concentric as 
if there had been an explosion in the center (see Fig.~\ref{fig:wfpc}).
There could be remnants of an additional, more diffuse and outer 
shell at a radius of $\sim 45\arcs$ from the nucleus; also  possibly
seen in the 8.0$\mu$m image south of the nucleus 
(see outer contours in Fig.~\ref{fig:wfpc}). 

The shell features are not coincident 
with the radio and X-ray knots seen 
by \citet{burns83,kraft2000}.
Viewing the Chandra observations, 
we did not see a strong excess of diffuse X-ray emission within the shell, 
however there may be a deficit of diffuse 
X-ray emission along the shell rim south-west of the nucleus 
(see Fig.~1 by \citealt{karovska02}).
No shell-like feature is seen in existing radio continuum images
of Centaurus~A (e.g., by \citealt{sarma02,clarke02}), 
however faint continuum emission associated
with the shell could be difficult to detect because
of the proximity of the bright jet and radio lobes.
As pointed out by \citet{marconi00}, there may be a reduction
in the extinction along the jet axis.  In the $8.0\mu$m image
there is a gap in the shell on the jet axis south-west of the nucleus.
North-west of the nucleus there is also a reduction
in mid-infrared flux in the shell rim along the jet axis, suggesting
that the jets have punctured holes in the shell. 

%Shells detected in other galaxies have primarily
%been detected in line emission from ionized gas
%and radio continuum (e.g., \citealt{jogee98,veilleux94}).
%It might be interesting to examine future higher angular resolution
%and more sensitive radio continuum images of this galaxy in search of a 
%counterpart for the bipolar shell.
%Star formation (as seen from the Pa$\alpha$ image) is 
%coincident with deep extinction features \citep{marconi00}
%and the parallelogram seen in the {\it Spitzer images} and so
%is primarily occurring in the warped dusty and gaseous disk.
%The shell features are not coincident with star clusters or HII regions 
%seen in the {WFPC2} and Pa$\alpha$ images.  Consequently the
%shell is not bright because it is coincident with 
%cites of recent star formation.

Recent spectroscopic studies have focused on the ionized emission
near the nucleus.  A diffuse, broad and blue-shifted spectral component 
at the nucleus was reported by \citet{marconi06}.  
It has a velocity $\sim 300$~km~s$^{-1}$ below the galaxy systemic velocity
and is fairly broad, with a width of 400~km~s$^{-1}$ \citep{marconi06}.
This component could be associated with the expanding shell.
Absorption lines in HI and CO have been seen against the
radio nucleus \citep{vanderhulst83,sarma02,eckart99,israel91,wiklind97}, 
however none of these are blue-shifted more than 20~km~s$^{-1}$ 
below the galaxy systemic velocity.  
We note that the HI and molecular band absorption spectra cited above
do not extend below $\sim 150$~km~s$^{-1}$ of the 
galaxy's systemic velocity so they
would have missed a broad or significantly blueshifted absorption
component.  
%A recent  Australia Telescope Compact Array (ATCA)
%spectrum taken by B\"arbel Koribalski failed to reveal
%a blueshifted HI absorption component between -1500 and 0~km~s$^{-1}$ of 
%the galaxy's systemic velocity with a column depth above  
%$N_H \sim 2 \times 10^{20}~{\rm cm}^{-2}$. 

%section 2.1
\subsection{Estimating the mass in the shell from the dust emission}

% par1
Here we follow the procedure 
previously carried out by \citet{bland03}
for estimating the mass in the shell at the Galactic center.
%An optically thin limb brightened shell is brighter at its edge
%than at its center.  
We estimate the column depth of the front of the shell from the 
surface brightness of the limb brightened edge.
The contrast between the peak surface brightness
at the edge compared to that of the front of the shell is given by
$C \approx 2 \left({r \over \delta}\right)^{1/2}$,
%in the case that the shell is resolved 
for a resolved shell,
where $r$ is the radius of the shell, and $\delta$ is its thickness   
\citep{bland03}.
The shell is resolved with a thickness of a few arcseconds and
radius $r \sim 30\arcs$ giving us an estimated contrast of $C \sim 6$.
From the background subtracted 
surface brightness of 10~MJy~sr$^{-1}$ at $8$ and $24\mu$m in the 
edge of the shell, we
estimate that the front of the shell, alone, would 
have a surface brightness of $\sim 2$~MJy~sr$^{-1}$ at the 
same wavelengths.

% par2
Because of the difficulty in subtracting the background 
emission, we cannot measure the infrared colors in the shell 
well enough to estimate a dust temperature.
We estimate the column depth in the shell from the flux
at $8$ and $24\mu$m using an estimated strength for the  
interstellar radiation field  which
sets the dust temperature in diffuse media \citep{lidraine01}.
%To estimate a column depth from the infrared peak surface brightness
%we must first estimate the strength of the interstellar radiation field.
Centaurus~A has a far infrared luminosity $\sim 10^{10} L_\odot$ 
\citep{eckart90} in
a region $\sim 10$kpc$^{-2}$, corresponding to a star formation
density in its disk of $\sim 0.1M_\odot~{\rm yr}^{-1}{\rm kpc}^{-2}$
(using conversion factors by \citealt{kennicutt98}).  This rate
is a few hundred times larger than that of the solar neighborhood
(estimated in the same way from the far infrared luminosity;
 \citealt{bronfman}).
We estimate that the ratio, $\chi$, of the UV radiation field 
to that of the Galactic UV interstellar radiation field 
in the solar neighborhood is a few hundred.
This level is consistent with the approximate ratio of 1 for 
the 8 and 24$\mu$m surface brightness in the shell.
Using column depths estimated for the diffuse ISM 
as a function of the interstellar radiation field \citep{lidraine01},
the mid-infrared surface brightness for the front
of the shell corresponds to a column depth of 
hydrogen of $N_H \sim 5 \times 10^{20}{\rm cm}^{-2} \chi_{100}^{-1}$.
Converting this to a mass we find  a mass in the shell of
$\sim 10^6 M_\odot \chi_{100}^{-1}$.
If the shell contains molecular material or if
the radiation field is dominated by the AGN,
then we may have underestimated the shell's mass.  
The estimated column depth implies that HI observations
covering a wider velocity range than existing observations
should be able to detect the shell 
in absorption against Centaurus~A's nucleus.

The shell is likely to be expanding.
Using the above estimated shell mass %of $10^6 M_\odot$,  
the shell would have kinetic energy
$E_{kin} \sim 10^{53} \chi_{100}^{-1} v_{100}^2 {\rm erg},$
where the expansion velocity $v_{100}$ is in units of 100~km~s$^{-1}$.
If the shell is moving faster than 100~km~s$^{-1}$,
then we would have underestimated its kinetic energy.
%
%{\bf  Timescale }
The sound speed of the ambient X-ray emitting material 
is $\sim 300$~km~s$^{-1}$.
The sound crossing time at 500~pc is $\sim $2~Myr.
We use this timescale as a possible estimate
for the age of the shell.

%section 2.2
\subsection{Energy estimates based on expansion of wind blown bubbles}

Wind blown bubble models predict the energy injection required
to create a bubble expanding into a uniform medium 
(e.g., \citealt{chevalier85,tomisaka88}).
If the energy injection is continuous, then
$dE/dt \sim 3 \times 10^{41} r_{kpc}^2 v_{100}^3 n_0~{\rm erg~s}^{-1}$
is required to produce a bubble of radius $r_{kpc}$, in units of kpc, 
velocity $v_{100}$ in units of 100~km~s$^{-1}$,
expanding into a medium with the ambient density, $n_0$, 
in units of cm$^{-3}$.
%and
%\begin{equation}
%dp/dt \sim 3 \times 10^{34} r_{kpc}^2 v_{100}^2 n_0~{\rm dynes}
%\end{equation}
%for momentum conservation 
Here we have used scaling laws given by \citet{castor75} 
for an energy conserving bubble.
Assuming a $\beta$ model with density $n(r) = n_0[1+ (r/a)^2]^{3\beta/2}$,
\citet{kraft03} estimate from the X-ray surface brightness distribution 
$n_0 \sim 0.04 cm^{-3}$ and $a=0.5$~kpc, though
the central density may be an underestimate because absorption from
the gaseous and dusty disk has not been taken into account.
The central region has an estimated density of $n_0$.
Inserting
$n_0=0.04$cm$^{-3}$ into the above scaling relation,
we find $dE/dt \sim 3 \times 10^{39} v_{100}^3 n_{0.04} {\rm erg~s}^{-1}$,
where $n_{0.04} = 0.04~{\rm cm}^{-3}$ is the density used.
Using the relation between mechanical luminosity and star formation
rate based on the population synthesis models by \citet{leitherer99},
this corresponds to a star formation rate of only $0.004 M_\odot$~yr$^{-1}$.
This power could easily be provided by the active nucleus which
has a bolometric luminosity of $10^{43}$ergs~s$^{-1}$ 
\citep{whysong04}.

If the injection was sudden then the total energy required to create
the bubble can be estimated using a Sedov-type expansion model.
In this case
$E_{kin} \sim 5 \times 10^{54} r_{kpc}^3 v_{100}^2 n_0~{\rm erg}$.
For our given radius and ambient density we 
estimate a total energy of 
$E_{kin} \sim 10^{52} v_{100}^2 n_{0.04}$~erg.
We note that this energy estimate is similar to that 
estimated above based on the dust mass.
This mechanical energy only requires a dozen supernovae or a total
mass of newly formed stars of order $\sim 1000 M_\odot$
(based on Figures 44 and 112, by \citealt{leitherer99}).
A higher ambient density or expansion velocity would lead
to increases in the required energy budget.
%We note that if the ambient density is an order of magnitude higher
%then an order of magnitude higher mechanical energy is required to create
%for the shell.  The shell velocity could also be significantly
%higher and this would also increase the

%\subsection{Comparison to estimates from other shells}

The size of the shell in Centaurus A is smaller than the shell or bubbles
of NGC~2992, NGC~4388, M82, NGC~3079 and NGC~2782 that are 
1 to a few kpc in size \citep{ulvestad84,kenney02,veilleux94,jogee98}.  
However Centaurus~A's shell is larger than the one
at the Galactic center which is only 170~pc \citep{bland03}.
Our energy estimates given above are highly uncertain because we
lack constraints on the shell's velocity.  If this shell is moving
at a 1000~km~s$^{-1}$ then it could have an energy as large as $10^{54}$~erg
and similar to that estimated for the Galactic center shell
or that present in NGC~3079.

%The kinetic energy
%estimated for NGC~4388 nuclear bipolar supershell 
%is $1.2 \times 10^{53}$ ergs \citep{kenney02}.
%This is similar to that in the post blow-out outflow of 
%M82 ($2 \times 10^{53}$ ergs) but lower than the early blowout phase
%outflow of
%NGC~3079 ($2 \times 10^{54} $ ergs \citep{veilleux94}.
%Velocites of up to 1100~km/s were measured in NGC~3079's outflow. This
%outflow has a radial extent of 1-2~kpc \citep{veilleux94}.
%
%The pre-blowout buble in NGC~2782 has a radial extent of 1~kpc.
%\citet{jogee98} estimate a timescale of $4\times 10^6$ and preblowout.
%\citep{jogee98} refers to velocites of a 100-300~km/s measured in this.
%The bipolar bubble at Galactic center is 170~pc big   
%and requires and estimated mechanical energy of 
%$\sim 10^{54}$erg \citep{bland03}.

\section{Summary and Discussion}

In this paper we have presented {\it Spitzer Space Telescope} 
observations of the nuclear region of the galaxy Centaurus A
that reveal a previously undetected 
500~pc radius shell above and below the gaseous and dusty
warped disk.  The shell resembles a bipolar outflow or bubble, 
has an axis ratio of $\sim 0.6$, 
a position angle of $\sim 10^\circ$ and a
surface brightness of  $\sim 10$~MJy~sr$^{-1}$ at 8.0$\mu$m.
It is extended in the direction perpendicular to the gaseous and dusty disk
and not in a direction obviously related to the radio jet.
%The shell is seen in three 
%bands and in both {IRAC} and {MIPS} instruments.
%Features in the shell are coincident with
%faint extinction features seen at visible bands.
We conservatively estimate that the shell contains
a million solar mass of hydrogen and that  
the energy required to create it is $\sim 10^{53}$ergs.   
Unfortunately we lack constraints on the shell's velocity.
If the shell is expanding at 1000~km~s$^{-1}$ then the energy required
could be 100 times larger.   While a blue shifted diffuse component
was detected at the nucleus by \citet{marconi06}, most spectroscopic
observations fail to cover the shell or lack the
bandwidth or sensitivity to have detected it.
Observational spectroscopic studies are needed to confirm the presence
of this transient shell and place better constraints on its mass, composition 
and energetics.

If the energy in the shell is low ($10^{53}$ ergs), then a modest
starburst of a few thousand solar masses 
could have provided the mechanical energy.
The orientation of the shell differs from
that of the radio jets so the shell may not have been caused
by the active nucleus.   However the bolometric luminosity
of the active nucleus exceeds that required to expand the bubble
so the active nucleus could have been important in its creation.
This shell is too small to 
have evacuated the 0.1 - 0.8 kpc gap in the dust 
distribution in the disk reported by \citet{quillen_irwarp},
suggesting that there could be 
or have been more than one expanding bubble in the heart of 
this galaxy.

\acknowledgments

%We are extremely grateful to B\"arbel Koribalski 
%for sending us an ACTA spectrum.
We thank Dan Watson, Bill Forrest, Eric Blackman, Todd Thompson, Daniel Stern,
%Jocelyn Keene, Charles Lawrence
and George Rieke for helpful suggestions and comments.
Support for this work was in part
provided by National Science Foundation grant AST-0406823,
and the National Aeronautics and Space Administration
under Grant No.~NNG04GM12G issued through
the Origins of Solar Systems Program.
We acknowledge support by award HST-GO-10173-09.A
through the Space Telescope Science Institute.
This work is based on observations made with the Spitzer Space Telescope,
operated by the Jet Propulsion
Laboratory, California Institute of Technology, under NASA contract 1407.

\clearpage

%Fig1

% Fig2

%Fig3

\end{document}